\documentclass[11pt,twoside]{article}

\usepackage{asp2006}
\usepackage{graphicx}

\begin{document}
\title{The Narrow Line Region:\\ Current Models and Future Questions}
\author{Brent Groves\altaffilmark{1}}
\affil{Max Planck Institute for Astrophysics, 
Karl-Schwarzschild Str. 1, 85741 Garching, Germany}
\altaffiltext{1}{Current Address: Leiden Observatory, Neils Bohr weg 2, 
Leiden 2333 CA, Netherlands; brent@strw.leidenuniv.nl}

\begin{abstract} 
I present a broad overview of modelling of the Narrow Line Region
(NLR) of active galaxies, and discuss some of the more recent models
we currently have for the emission from the NLR. I show
why the emission line ratios from the NLR are constrained to
certain observed values, and describe what physical parameters we can
derive from observations using emission line models. Also presented are some
examples of this, looking at the metallicity and excitation mechanism
of active galaxies. As a final point, the limitations of the
current models are discussed, and how how the combination of modelling and theory can
help us solve some of the questions that still remain within the NLR.
\end{abstract}

\section{Introduction}

The Narrow Line Region (NLR) is the region of extended interstellar gas
ionized and heated by the active galactic nucleus (AGN). This region
is classified as ``narrow-line'' as the gas lies outside the
dominating influence of the central black hole, and the observed line
velocity widths are much 
less than found in typical Broad Line Regions (discussed earlier in these
proceedings). Even so, typical NLR velocity widths are in the range
$200<$FWHM$<500$ km s$^{-1}$, and even higher velocity tails are
observed.

Another distinguishing feature of the NLR is the
gas density. In the NLR, the gas density is low enough that emission
lines arising from magnetic dipole transitions can occur. These
forbidden lines, like [O\,{\sc iii}]$\lambda 5007$\AA\ and [N\,{\sc
ii}]$\lambda 6584$\AA, are strong gas
coolants and dominate the NLR emission line spectrum. It is these
lines that are typically used to identify both the NLR and AGN in
optical spectra.

The NLR is also much more spatially extended than the BLR, reaching kiloparsec
scales if the ``Extended Narrow Line Region'' (ENLR) is included. 
At such distances the NLR extends well beyond the nuclear
obscuring material commonly called the torus, meaning that it is generally still
visible even when other more nuclear regions, like the BLR, are hidden.

\section{Line Ratio Diagrams}

In modelling the NLR, we want to reproduce and explain the geometry and spatial
extent of the region and, most importantly, the emission from this
region. The emission lines from the NLR are not only
indicative of the region, but also provide diagnostics for the
density, temperature, and ionization mechanism of the NLR gas.

\begin{figure}
\includegraphics[width=\hsize]{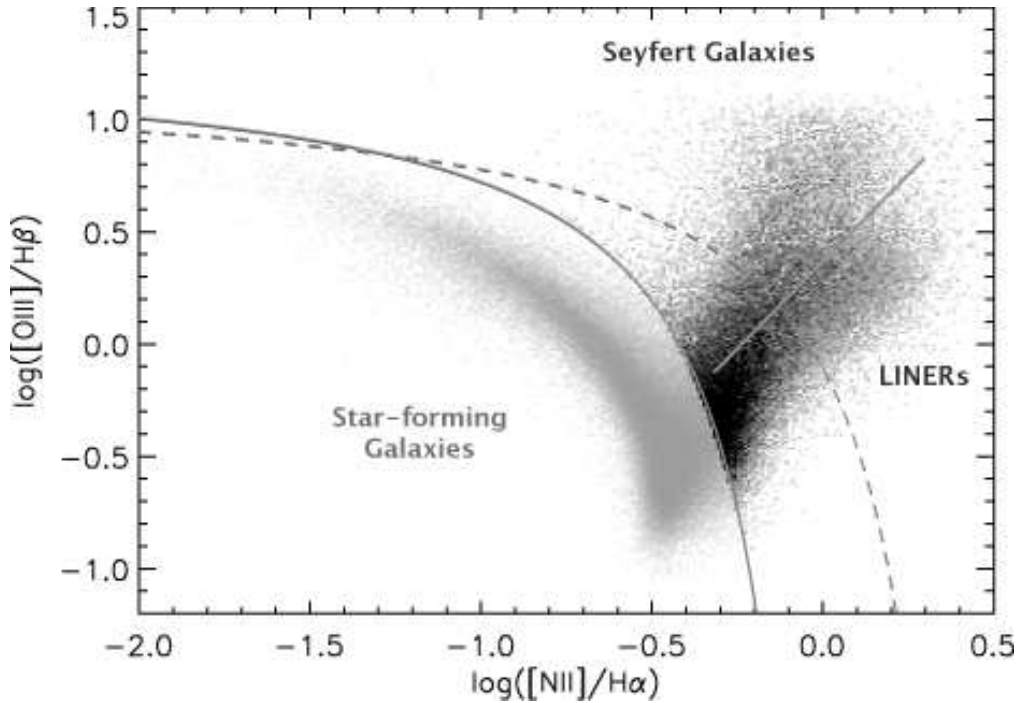}
\caption{BPT diagram of [O\,{\sc iii}]$\lambda 5007$/H$\beta$ vs.~[N\,{\sc
ii}]$\lambda 6584$/H$\alpha$, showing the distribution of SDSS DR4
emission line galaxies. The three classes of emission line galaxies,
Star-forming, LINERs and Seyferts, are seen as three separate
branches, with the empirical divisions marked to guide the eye. The
LINERs and Seyferts are often classified together as an ``AGN
branch'' \citep[\emph{Figure 1} from ]{GrovesZ06}.}\label{fig:BPTex}
\end{figure}

A useful visual tool for comparing models with observations of many
NLR are Line Ratio Diagrams. As demonstrated by figure \ref{fig:BPTex}, these
plot ratios of emission lines against each other and, depending upon
the emission lines chosen, can show clear
relationships with density, metallicity, and ionization mechanism
\citep{Veilleux87}. Figure \ref{fig:BPTex} shows the ``BPT'' diagram,
one of the earliest and strongest emission line diagrams, suggested by
\citet{Baldwin81}. This diagram is able to distinguish three different
classes of narrow emission line galaxies: star-forming galaxies, Low
Ionization Narrow Emission line Region or LINER galaxies, and the Seyfert
galaxies which show typical AGN NLR emission line ratios. These
three classes differ in their ionizing source and hence nebular
temperature and emission. In terms of the NLR models, the final
reproduced spectra must lie in the top right section of this diagram
in the region occupied by the ``pure'' Seyfert galaxies.

\section{Emission Line Region Modelling}

To determine the resulting emission lines from ionized nebulae,
emission line region models need to determine the density, temperature
and ionization state of the ionized gas \citep[discussed in detail
in][]{ADU03}. The temperature and ionization are determined
fundamentally by the ionization source, of which there are two main
pathways: Photoionization and Shock ionization.

\subsection{Shock Ionization}

In shock ionization modelling of NLR, the gas is excited collisionally
through shocks caused by interactions with a jet or winds arising from
the central AGN source. These models generally take as input the gas
abundances and pre-shock density, shock velocity $V_{s}$, and a
parameter related to the magnetic field strength $B/n^{1/2}$. 

\begin{figure}
\center
\includegraphics[width=0.8\hsize]{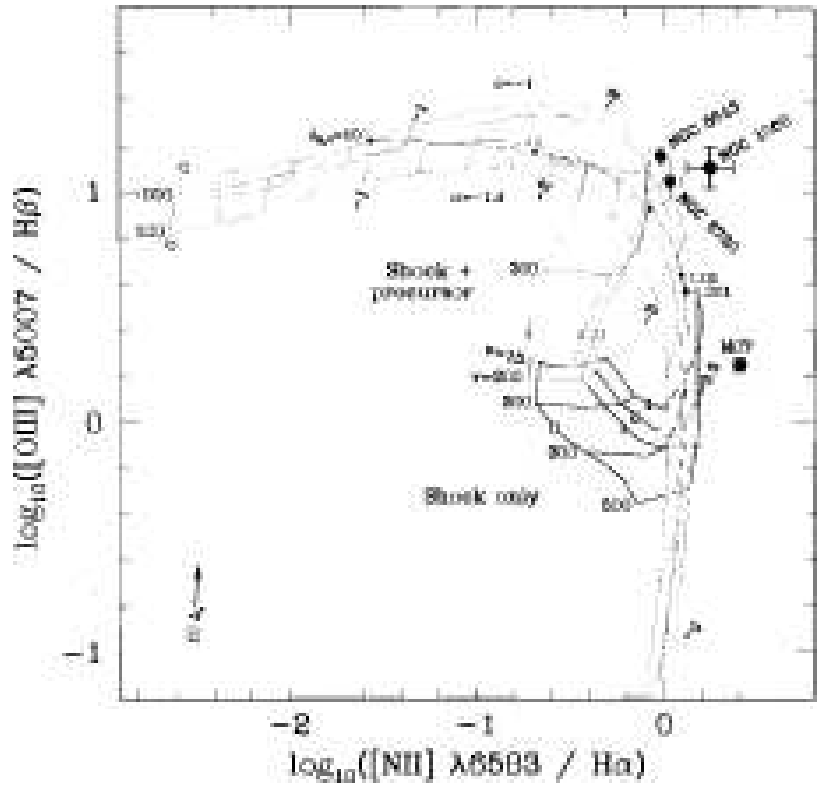}
\caption{BPT diagram showing four different models as labelled. Four
nearby AGN are marked; 3 typical Seyferts and M87, known to have a
strong jet and shocks. \citep[\emph{Figure 1} from ][]{Allen98}}
\label{fig:BPTmodels}
\end{figure}

Typical line values for shock models are shown in figure
\ref{fig:BPTmodels}. These models, discussed in detail in
\citet{Allen98}, cover the range of parameter space expected for
Seyfert NLRs. As can be seen from the figure, the simple shock models are
unable to reproduce typical Seyfert NLR line ratios. For some LINER-like
objects however, they might possibly be the exciting mechanism.

\subsection{Photoionization}

In photoionization modelling, the NLR is excited by the UV and X-ray
photons from a central source, thought to be the accretion disk
surrounding the central black hole in AGN. These models have been
explored for many years, with good reviews being the book by
\citet{Osterbrock89} and the lecture by H. Netzer \citep[in ]{AGN90}.

The typical parameters for photoionization models are the gas
abundances and initial density, the size or column depth of the model
cloud, the incident ionizing spectrum, and the incident ionizing flux
of the radiation. The final parameter is often described in terms of
the ionization parameter, $U$, a dimensionless measure of the density
of ionizing photons over the gas density,
\begin{equation}
U=\frac{1}{n_{\rm H}c}\int^{\infty}_{13.6~{\rm eV}}\frac{F_{\nu}}{h\nu}
d\nu.
\end{equation}
$U$ is  a controlling parameter of the emission line spectrum in
photoionization models.

The resulting models for a range of typical NLR values can be seen
in figure \ref{fig:BPTmodels}. The models do reproduce the ratios
quite well for a limited range of parameters, as well as others such
as [O\,{\sc I}]$\lambda 6300$/H$\alpha$. However these models cannot
reproduce both low-ionization and high-ionization line strengths
simultaneously, for example failing to reproduce He\,{\sc II}$\lambda
4686$/H$\beta$ with the same parameter set, and these simple models
haven been ruled out \citep{Stasinska84}.

With both simple shock models and simple photoionization models not
surprisingly ruled out, several more complicated models have been put
forward to try and better reproduce and explain the NLR.

\subsection{Shock+Precursor}

It was realized early on in shock modelling that fast shocks ($V_{s}>150$ km
s$^{-1}$) would produce ionizing
photons \citep{Dopita95,Dopita96}. As post-shock gas cools it produces ionizing photons
which diffuse up- and down-stream, and ionize the pre-shock
gas. Thus the ionization in fast shocks is actually a combination of both shock and
photo-ionization, where the photoionization is determined by the
shock velocity.

These shock+precursor models are able to reproduce the observations
quite well, as shown in figure \ref{fig:BPTmodels}, and can also
produce some of the higher ionization lines. However, the problem with
these models is that they require shocks throughout the NLR, meaning 
shock signatures should always be visible. So by themselves these models
cannot explain all NLR emission.

\subsection{Multi-Component Photoionization}

The next level of complexity in terms of photoionization models is
multi-component or multi-cloud photoionization, where the combination
of two or more photoionization models is used to reproduce both the high
and low ionization lines in the NLR.

Generally, most models limit themselves to two components to minimize
the number of free parameters. These models have been used to examine
specific galaxies \citep[e.g.~][]{Kraemer00,Morganti91} and more generally
trying to explain specific line ratios
\citep[e.g.~][]{Komossa97,Murayama98} or line strengths \citep{Baskin05}.

The main problem with these models is that as you increase the number
of clouds, the problem becomes unconstrained. A way around this is to
provide a physical basis for the multiple clouds, and hence physical
constraints. Here I mention three
current models that try to deal with this problem.

\subsection{$A_{\rm M/I}$ Models}

The $A_{\rm M/I}$ models of \citet{bws96} consider the NLR to be made
up of hot, highly ionized, matter bounded\footnote{Matter bounded
clouds have low enough column densities that not all ionizing
radiation is absorbed. This compares to Ionization bounded clouds
where all ionizing radiation is absorbed.} clouds and, lower
ionization, ionization bounded clouds. In their model, the ionization
bounded (I-) clouds see the absorbed spectrum from the matter bounded
(M-) clouds, and the I-clouds are also a higher density. The resulting
emission line spectrum from these models is the controlled by the
ratio of these two components, $A_{\rm M/I}$. 

\begin{figure}
\center
\includegraphics[width=0.8\hsize]{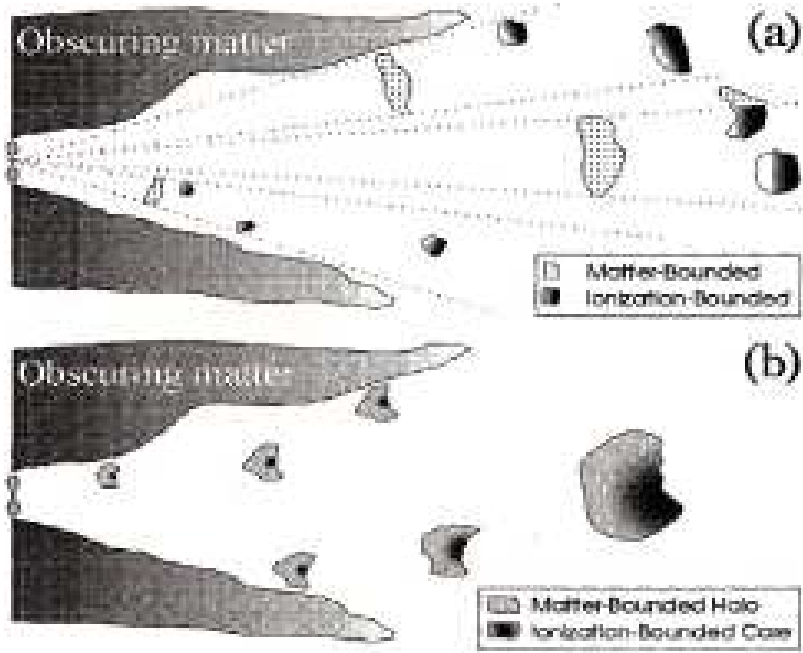}
\caption{Two possible physical geometries for the $A_{\rm M/I}$
model. \citep[\emph{Figure 4} from ][]{bws96}} 
\label{fig:AMI}
\end{figure}

Two possible physical pictures of their model are shown in figure \ref{fig:AMI}
(their figure 4). The first (figure \ref{fig:AMI}a) is where the
clouds are separate, with inner clouds being M-clouds, and the outer
are I-clouds, while the second (figure \ref{fig:AMI}b) shows
composite clouds, with an I-cloud core and a M-cloud photoevaporating halo.

These models reproduce the strong lines quite well as shown in figure
\ref{fig:BPTmodels} by the $A_{\rm M/I}$ marked curve, and they also
reproduce very well strong high ionization ratios such as He\,{\sc II}$\lambda
4686$/H$\beta$ \citep[see e.g.~fig. 7,][]{bws96}

\subsection{Local Optimally Emitting Clouds}

The Locally Optimally-emitting Cloud model of \citet{Ferg97} is an
extension of a model for the BLR. It uses the fact that the spectra we
observe in AGN NLRs are likely to be dominated by selection effects:
each line we observed arises from the clouds in which it is most
strongly emitted. As each line emits strongest near its critical
density, this recreates the linewidth-critical density relation
observed in some NLR.

For their model, \citeauthor{Ferg97} run a grid of simple
photoionization models covering a range of densities and radii from
the nucleus (connected to incident flux). The total line flux is then
the integral of these over the NLR cloud distribution function, $\psi$,
\begin{equation}
F_{\rm line}\propto\int\int r^2 F_{\rm model}(r,n)\psi(r,n)dn~dr.
\end{equation}
While $\psi(r,n)$ can take any form, for simplicity they assume
separate power-law distributions,$\psi(r,n)\propto
r^{\alpha}n^{\beta}$. With this distribution and simple models they
obtain models that reproduce the observations very well, as shown by
their figure. However
again, these models are actually relatively physically unconstrained,
and importantly they ignore the fact that the NLR is clumpy in nature
and dusty.

\subsection{Radiation Pressure Dominated Dusty Clouds}

The dusty, radiation pressure dominated cloud models by
\citet{Groves04a,Groves04b} take in the observations that indicate
that the NLR clouds are likely dusty and clumpy in nature to come up with a
physical model that explains why AGN NLRs cluster where they do on
line ratio diagrams. It uses the fact that at high ionization
parameter ($U$) dust dominates the opacity in dusty gas, and that in
isobaric systems the gas pressure gradient must match the radation
pressure gradient to realise that radiation pressure on dust will
dominate the NLR structure at high $U$. As shown in detail in
\citet{Dopita02} this leads to a self regulatory mechanism for the
{\bf local}\footnote{$U_{\rm local}$ is defined using the local
absorbed ration field and density.} ionization parameter and hence for
the emission lines. These models reproduce the observations very well
for both high and low ionization species, as shown in the figures in
\citet{Groves04b}, and converge in the region in line ratio
diagrams where the NLR is observed to be. 

\begin{figure}
\center
\includegraphics[width=0.9\hsize]{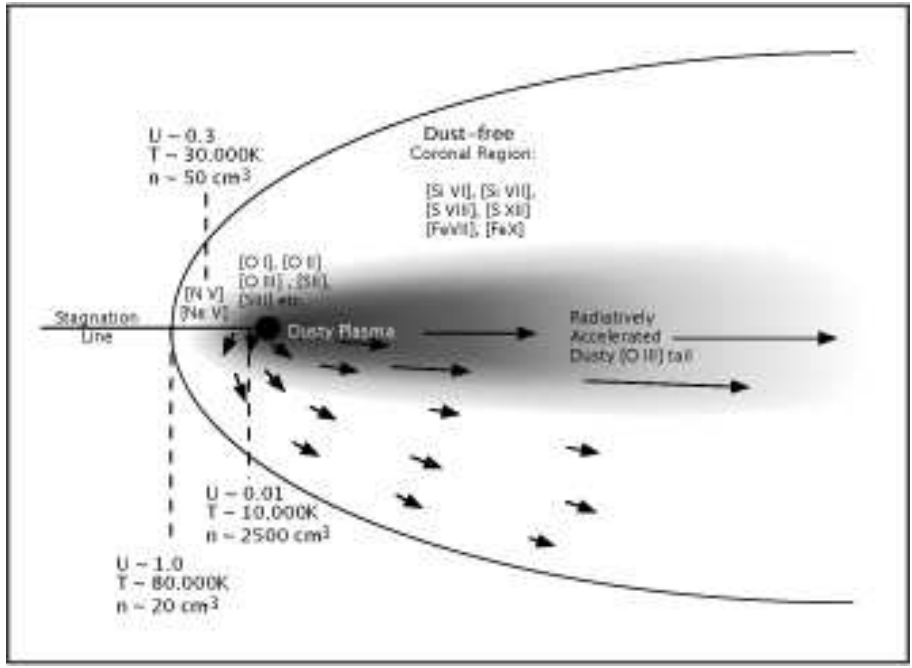}
\caption{Possible physical picture for a NLR cloud in the Dusty,
P$_{\rm rad}$ model. \citep[\emph{Figure 1} from ][]{Dopita02}} 
\label{fig:NLRmodel}
\end{figure}

The physical picture for these models is shown in figure
\ref{fig:NLRmodel}, revealing the main emitting region, the observed
NLR tails, and possible sources for the very high ionization ``coronal
lines'' observed in some AGN. In many respects it is very similar to the 
$A_{\rm M/I}$ picture shown in figure \ref{fig:AMI}b.

These models have several other benefits arising from the dust, such as a
hardening of the radiation field (cf.~the $A_{\rm M/I}$ models),
increased temperature through photoelectric heating, and of course the
corresponding IR emission. This provides another constraint of NLR
models, as the models \citep[e.g.~][]{GrovesIR06} must match recent
observations of IR emission from the NLR (see
e.g.~\citet{Bock00,Mason06} and several other papers in this
proceedings such as by Schweitzer).  

The dust is also a hindrance in some respects, as Iron lines like
[Fe\,{\sc VII}]  
tend to be too weak due to dust depletion. Similarly, other high
ionization coronal lines are unable to be reproduced. It should be
noted however, that these models are for individual NLR clouds, not the NLR
as a whole.

\section{Low Metallicity AGN}

\begin{figure}
\includegraphics[width=\hsize]{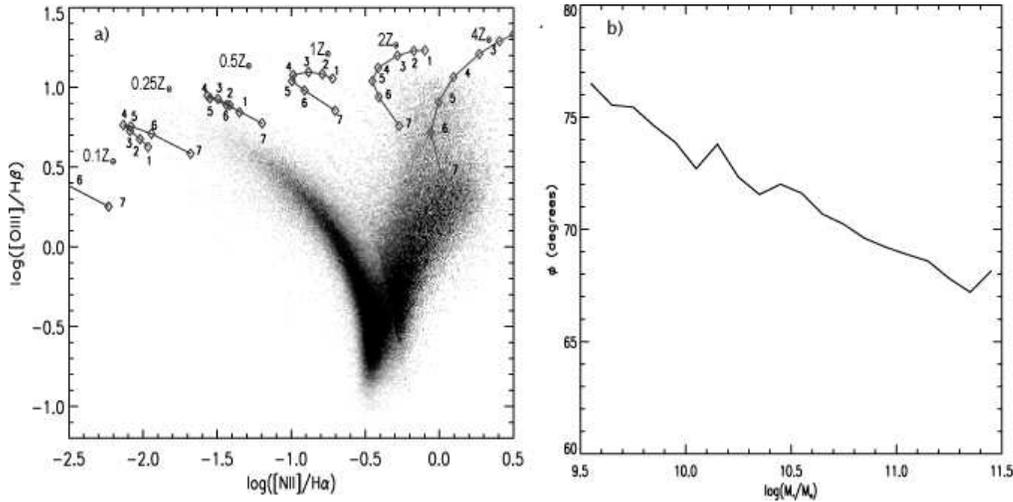}
\caption{The effects of low metallicity on AGN. Fig. a) shows NLR models of
decreasing metallicity (as labelled) on the BPT diagram. The
underlying image shows the distribution of the SDSS DR4 emission line
galaxy sample. Fig. b) shows how the mass-metallicity correlation
applies in AGN as well, showing how the galaxy mass varies with
position in the BPT diagram (as defined in \citet{GrovesZ06}) which
correlates with metallicity. \citep[Figures from ][]{GrovesZ06}.} 
\label{fig:lowZagn}
\end{figure}

One of the uses of NLR models is to determine the average physical
quantities of a specific AGN. Inversely, it is possible to use NLR
models to predict what type of emission AGN of a specific type would
have. For example, in a recent work \citet{GrovesZ06} used NLR models
to determine strong metallicity-sensitive emission line ratio
diagrams, and where low metallicity AGN would appear on these.
Figure \ref{fig:lowZagn}a reveals one possible diagram, the BPT
diagram, well known to be metallicity sensitive due to the secondary
nature of nitrogen \citep{Storchi98}. These diagrams were then used to
look for possible low metallicity AGN in the SDSS emission line galaxy
sample. Only 40 likely low metallicity Seyferts were found in a sample
of  $\sim$23000 Seyferts (or $\sim 8800$ ``pure'' Seyferts).

In figure \ref{fig:lowZagn}b) the variation of the galaxy host mass
with position in the BPT Seyfert branch is shown
\citep[details found in][]{GrovesZ06}. This reveals that
the mass-metallicity correlation holds in AGN as well (as expected). 

\section{Diagnosing Emission Line Galaxies}

\begin{figure}
\centering
\includegraphics[width=0.9\hsize]{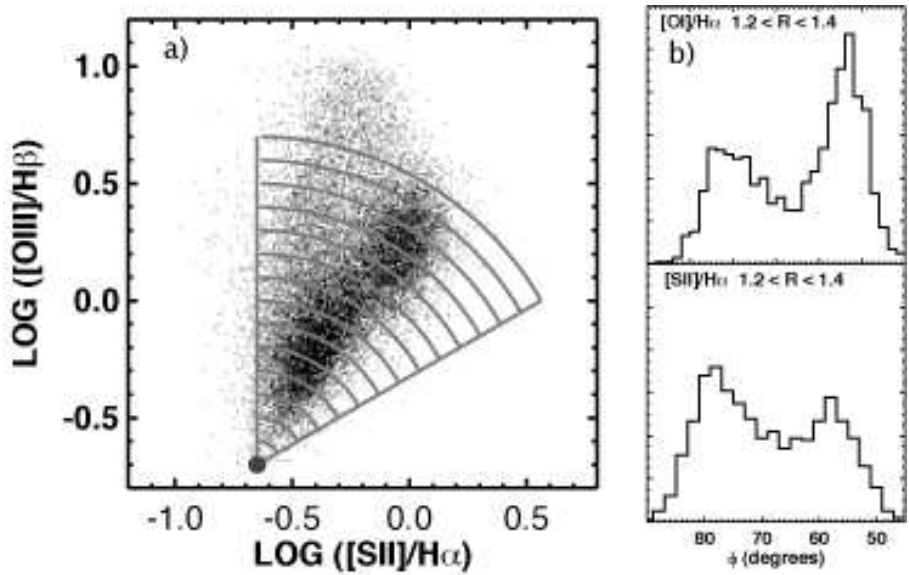}
\caption{The separation of LINERs from Seyferts. Figure a) shows the
AGN branch of the SDSS galaxies on the [S\,{\sc
II}]$\lambda\lambda6713,31$ line ratio diagram, and the placement of
the arcs to measure the separation of the Seyfert and LINER
branches. Figure b) shows the clear separation of the Seyfert and
LINER branches in the [S\,{\sc II}]$\lambda\lambda6713,31$ and
[O\,{\sc I}]$\lambda 6300$ ratio diagrams. }
\label{fig:Sy_L}
\end{figure}

Emission line ratio diagrams were originally suggested to separate
Star-forming galaxies from Seyfert type-2 (NLR only) galaxies, as both
show strong narrow lines, although at different strengths
\citep{Baldwin81}.  With the advent of large spectroscopic surveys
like SDSS this work has progressed much further, with the BPT diagram
showing a clear continuum of objects with the edges of the V-Shaped
figure corresponding to galaxy evolution effects, ionization limits,
and observation limits. While the distinction between the star-forming
galaxies and AGN is clear in the BPT diagram (Figure \ref{fig:BPTex})
apart from at low [O\,{\sc III}]/H$\beta$, the distinction
between the LINERs and Seyfert galaxies is not so clear, and these
objects are often classed together. Recent work
by \citet{Kewley06} however, has shown that a clear divide does exist
between Seyfert and LINER objects as shown in figure
\ref{fig:Sy_L}. This figure demonstrates that LINERs and Seyferts
appear as distinct populations in the [S\,{\sc II}] and [O\,{\sc I}]
line ratio diagrams.

With this distinction, the separate galaxy properties of all three
classes have been examined and compared. In addition, with NLR models like
\citet{Groves04a,Groves04b} and Star-forming galaxy models like
\citet{Dopita06} we are coming a far way in understanding the full
spread of emission line galaxies in line ratio diagrams.

\section{Conclusions}

With current Narrow Line Region models we have a reasonably good
physical understanding of the 
NLR and its appearance and emission lines. These models can be used to
determine properties of the NLR and AGN, but caution is still needed,
as these models are limited. 

An individual NLR consists of many
clouds, with a probable range of densities, pressures and incident
spectrum. These clouds could also vary in their abundances,
metallicity and dust properties, and certainly vary in shape and
size. The differences between different active galaxies will be even
larger. Thus models must account for this, but similarly they must
account for the fact that NLR spectra are all very similar. The current
NLR models are almost there but a full physical, geometrical model is
not yet available. 

\acknowledgements 
I would like to thank the organizers for inviting me to what turned
out to be a full 
and very interesting conference, and for giving me the opportunity to visit China.
I would also like to thank MPA and G. Kauffmann for funding me to
this conference.


\end{document}